\documentclass[aip,jcp,reprint,floatfix]{revtex4-1}
\usepackage{graphicx}
\usepackage{dcolumn}
\usepackage{bm}
\usepackage[utf8]{inputenc}
\usepackage[T1]{fontenc}
\usepackage{mathptmx}
\usepackage{etoolbox}
\usepackage{balance}
\usepackage{xcolor}
\usepackage{soul}
\usepackage{amsmath}
\usepackage{braket}
\usepackage[version=4]{mhchem}
\usepackage{physics}
\usepackage{physics2}
\usepackage{booktabs}
\usepackage{array}
\usepackage{multirow}
\usepackage{tabularx}
\usepackage{threeparttable}
\usepackage{adjustbox}
\newcommand{\Lower}[1]{\smash{\lower 1.5ex \hbox{#1}}}
\newcommand{\Tp}{\hat T_{\rm p}}

\newcommand{\Tt}{\hat T}
\newcommand{\Hn}{\hat{\mathcal{H}}_N^{(\rm CC)}}
\newcommand{\Hp}{\hat{\mathcal{H}}_N^{(\rm pCCD)}}

\newcommand{\Hfp}{\hat{\mathcal{H}}_N^{(\rm fpCC)}}

\makeatletter
\def\@email#1#2{%
  \endgroup
  \patchcmd{\titleblock@produce}
    {\frontmatter@RRAPformat}
    {\frontmatter@RRAPformat{\produce@RRAP{*#1\href{mailto:#2}{#2}}}\frontmatter@RRAPformat}
    {}{}
}
\makeatother

\begin{document}

\preprint{AIP/123-QED}

\title[Excited state properties from Equation-of-Motion Frozen Pair Coupled Cluster Methods]
{Oscillator Strengths and Transition Dipole Moments from a simplified Equation-of-Motion Coupled Cluster Formalism within the Frozen-Pair Approximation}

\author{Seyedehdelaram Jahani}
\affiliation{Institute of Physics, Faculty of Physics, Astronomy, and Informatics, Nicolaus Copernicus University in Toruń, Grudziądzka 5, 87-100 Toruń, Poland.}

\author{Katharina Boguslawski}
\affiliation{Institute of Physics, Faculty of Physics, Astronomy, and Informatics, Nicolaus Copernicus University in Toruń, Grudziądzka 5, 87-100 Toruń, Poland.}
\email{k.boguslawski@umk.pl}

\author{Paweł Tecmer}
\affiliation{Institute of Physics, Faculty of Physics, Astronomy, and Informatics, Nicolaus Copernicus University in Toruń, Grudziądzka 5, 87-100 Toruń, Poland.}
\email{ptecmer@umk.pl}

\date{\today}
%%%%%%%%%%%%%%%%%%%%%%%%%%%%%%%%%%%%%%%%%%%%%%%%%%%%
\begin{abstract}
%%%%%%%%%%%%%%%%%%%%%%%%%%%%%%%%%%%%%%%%%%%%%%%%%%%%
In this work, we derive the working equations for the transition density matrices within the equation-of-motion frozen-pair coupled-cluster framework.
We focus specifically on the recently developed EOM-fpCCSD and EOM-ptCCSD models and present the corresponding transition dipole moments and oscillator strengths obtained via approximate expressions.
This formulation avoids the computational overhead of solving the coupled-cluster $\Lambda$ equations.
Furthermore, we employ a matrix inverse approximation to eliminate the need for explicit calculation of the left EOM eigenvectors.
The accuracy of the resulting EOM-fpCCD and EOM-ptCCSD excited-state properties is benchmarked against the linear-response (LR)-CCSD method. 
Our results demonstrate that the description of excited state properties is improved when using EOM-fpCCSD and EOM-ptCCSD models compared to the standard EOM-CCSD variant. 

%%%%%%%%%%%%%%%%%%%%%%%%%%%%%%%%%%%%%%%%%%%%%%%%%%%%
\end{abstract}
%%%%%%%%%%%%%%%%%%%%%%%%%%%%%%%%%%%%%%%%%%%%%%%%%%%%
\maketitle
%%%%%%%%%%%%%%%%%%%%%%%%%%%%%%%%%%%%%%%%%%%%%%%%%%%%
%%%%%%        Introduction
%%%%%%%%%%%%%%%%%%%%%%%%%%%%%%%%%%%%%%%%%%%%%%%%%%%%
\section{Introduction} \label{sec:introduction}
%%%%%%%%%%%%%%%%%%%%%%%%%%%%%%%%%%%%%%%%%%%%%%%%%%%%
%%%%%%%%%%%%%%%%%%%%%%%%%%%%%%%%%%%%%%%%%%%%%%%%%%%%
A reliable description of electronic excited states and their corresponding properties, such as oscillator strengths (OSs) and transition dipole moments (TDMs), of large molecules remains a real challenge for present-day quantum-chemical approaches.~\cite{td-dft-vs-eom-ccsd-os-benchmark-jctc-2011, quest-jctc-2020, quest-jctc-2021,pccd-koopmans-delaram-jcp-2025,ip-ekt-delaram-jctc-2026}
Despite recent methodological and computer hardware advances, robust and accurate wave-function-based methods remain computationally too expensive to routinely guide the experimental synthesis of organic electronic materials, predict their UV-Vis spectra, or model realistic organic devices.~\cite{opv-cui-jmcc-2020,risko-opv-cr-2023}
While time-dependent density functional theory (TD-DFT) offers a computationally affordable alternative, it frequently suffers from strong sensitivity to the choice of exchange-correlation functional, incorrect bond alternation in extended conjugated systems, and significant difficulties in accurately describing charge-transfer states, Rydberg transitions, and states with substantial double-excitation character.~\cite{dft-tddft-failure-extended-systems-jcp-2002, tortorella-dft-failure-osc-jpcm-2016}
Consequently, there is a strong need for methods that offer a favorable trade-off between computational cost and accuracy—retaining the high accuracy and systematic improvability of coupled-cluster theory while significantly reducing its steep computational scaling.
Several coupled-cluster-based approaches have been developed to address this challenge.~\cite{multi-level-cc-jcp-2014, tdm-xcc-jcp-2014, tdm-xcc-jcp-2017, eom-vertical-ee--dutta-jctc-2018, mlccsd-core-excitations-jctc-2020,ascc-jctc-2024, ascc-jpcl-2025, ascc-jpcl-2025, block-correlated-gvb-eom-jcpl-2025, eom-four-block-correlated-cc-jpcl-2026}
Promising are the frozen-pair coupled-cluster methods~\cite{tailored-cc-oliphant-jcp-1991, tailored-cc-implemntation-oliphant-jcp-1992, tailored-cc-jcp-1993, tailored-cc-jcp-1994, tailored-cc-mp-1998, ap1rog-lcc-jctc-2015, pccd-tailoed-drmg-jctc-2021, ea-eom-fpccd-jctc-2025} built upon the pair Coupled Cluster Doubles (pCCD) ansatz,~\cite{limacher-ap1rog-jctc-2013,oo-ap1rog-prb-2014, tamar-pccd-jcp-2014, ap1rog-non-variational-orbital-optimizarion-jctc-2014, geminal-review-pccp-2022} extended to excited state calculations~\cite{eom-pccd-jcp-2016, eom-pccd-erratum-jcp-2017, eom-pccd-lccsd-jctc-2019,pccd-delaram-rsc-adv-2023} such as Equation-of-Motion frozen-pair (EOM-fp)CCSD~\cite{eom-fpccsd-faraday-disccus-2026} and EOM-pair-tailored (pt)CCSD.~\cite{pccd-tailored-bartlett-jcp-2023}
These methods have emerged as attractive alternatives to conventional EOM-CCSD.~\cite{eom-fpccsd-faraday-disccus-2026}
Recent studies have shown that EOM-fpCCSD provides excitation energies for charge-transfer (CT) states that closely match those obtained with EOM-CCSD, while outperforming the EOM-ptCCSD variant.
Yet, EOM-fpCCSD offers a significantly improved description of doubly excited states and achieves more robust convergence in excited-state calculations for complex molecular systems compared with both EOM-ptCCSD and conventional EOM-CCSD.

These findings motivate us to further investigate the EOM-fpCCSD and EOM-ptCCSD models beyond excitation energies, with particular emphasis on their excited-state properties, such as TDMs and OSs.
That allows us to examine how these properties depend on the choice of orbital basis by comparing canonical Hartree--Fock (HF) and natural pCCD orbitals.
To achieve this at low computational cost, we follow the strategy proposed by Achintya et al.~[\citenum{eom-ccsd-expectation-value-tdms-jcp-2017}] and combine approximate EOM-CC expressions for the transition density matrix~\cite{xcc-ijqc-1993, td-xcc-collect-czech-chem-commun-2005, korona-s-operator-cc-jcp-2006, tdm-xcc-jcp-2014, tdm-xcc-jcp-2017, pccd-expectation-value-1dm-jpca-2025} with a matrix-inverse approximation.~\cite{eom-ccsd-expectation-value-tdms-jcp-2017}
We benchmark the newly developed EOM-fpCCSD and EOM-ptCCSD models against the linear-response (LR)-CCSD method for small molecules using a range of basis sets. 
%We then demonstrate the practical utility and predictive power of our approach on larger, more complex molecular systems.
%%%%%%%%%%%%%%%%%%%%%%%%%%%%%%%%%%%%%%%%%%%%%%%%%%%%
%%%%%%        Theory
%%%%%%%%%%%%%%%%%%%%%%%%%%%%%%%%%%%%%%%%%%%%%%%%%%%%
\section{Theory}\label{sec:theory}
%%%%%%%%%%%%%%%%%%%%%%%%%%%%%%%%%%%%%%%%%%%%%%%%%%%%
%%%%%%%%%%%%%%%%%%%%%%%%%%%%%%%%%%%%%%%%%%%%%%%%%%%%
\subsection{Ground-state electronic structures}
%%%%%%%%%%%%%%%%%%%%%%%%%%%%%%%%%%%%%%%%%%%%%%%%%%%%
%%%%%%%%%%%%%%%%%%%%%%%%%%%%%%%%%%%%%%%%%%%%%%%%%%%%
Our starting point for the EOM-fpCCD and EOM-ptCCSD methods is the pCCD ground-state wavefunction,~\cite{limacher-ap1rog-jctc-2013, tamar-pccd-jcp-2014, geminal-review-pccp-2022}
%%%%%%%%%%%%%%%%%%%%%%%%%%%%%%%%%%%%%%%%%%%%%%%%%%%%%%%%%%%%%%%%%%%%%%
%%%%%%%%%%%%%%%%%%%%%%%%%%%%%%%%%%%%%%%%%%%%%%%%%%%%%%%%%%%%%%%%%%%%%%
\begin{equation}\label{eq:pccd}
\ket{\rm pCCD} = \exp \left (  \sum_{i=1}^{\rm occ} \sum_{a=1}^{\rm virt} t_{i\bar{i}}^{a\bar{a}} 
    \hat a_a^{\dagger}  \hat a_{\bar{a}}^{\dagger}\hat a_{\bar{i}} \hat a_{i}  \right ) \ket{\Phi_0} = e^{\Tp} \ket{\Phi_0},
\end{equation}
%%%%%%%%%%%%%%%%%%%%%%%%%%%%%%%%%%%%%%%%%%%%%%%%%%%%%%%%%%%%%%%%%%%%%%
%%%%%%%%%%%%%%%%%%%%%%%%%%%%%%%%%%%%%%%%%%%%%%%%%%%%%%%%%%%%%%%%%%%%%%
where $\Tp$ is the electron-pair excitation operator, containing electron creation ($\hat a^\dagger_p$ for $\alpha$ spin and $\hat a^\dagger_{\bar{p}}$ for $\beta$ spin) and annihilation ($\hat a_p$ for $\alpha$ spin and $\hat a_{\bar{p}}$ for $\beta$ spin) operators and the electron-pair cluster amplitudes $t_{i\bar{i}}^{a\bar{a}}$.
In the above equation, the sum runs over all occupied $i$ and virtual $a$ orbitals of this reference state and $| \Phi_0 \rangle$ is some determinant. 
The reference determinant and its orbitals are usually optimized variationally using a Lagrangian formulation.~\cite{oo-ap1rog-prb-2014,tamar-pccd-jcp-2014,ap1rog-manual-orbital-rotations-mp-2014,ps2-ap1rog-jcp-2014,ap1rog-non-variational-orbital-optimizarion-jctc-2014}
Orbital optimization yields natural pCCD orbitals that are localized, allowing us to model strongly correlated systems.~\cite{diatomics-oo-ap1rog-jpca_2014, ap1rog-singlet-gs-actinides-pccp-2015,pccd-correlation-analysis-prb-2016, state-specific-oopccd-jctc-2021}  
By construction, pCCD describes electron correlation effects restricted to electron pairs, and the cluster operator spans only the seniority-zero sector of the wavefunction expansion.~\cite{limacher-ap1rog-jctc-2013,diatomics-oo-ap1rog-jpca_2014,tamar-pccd-jcp-2014, oo-ap1rog-prb-2014, ps2-ap1rog-jcp-2014, ap1rog-manual-orbital-rotations-mp-2014, seniority-zero-mean-field-wfn-jcp-2025}

The missing broken-pair states in the pCCD model, can be accounted for using, for instance, perturbation theory~\cite{ap1rog-pt2b-pccp-2014,pccd-ptx-jctc-2017, seniority-0-canonical-transformation-theory-jcp-2026} or CC corrections.~\cite{frozen-pccd-jcp-2014,ap1rog-lcc-jctc-2015,pccd-tailored-jctc-2022}
Here, we focus on the latter. 
CC corrections on top of pCCD can be understood in terms of tailored Coupled Cluster (tCC) theory,~\cite{tailored-cc-oliphant-jcp-1991,tailored-cc-implemntation-oliphant-jcp-1992,tailored-cc-jcp-1993,tailored-cc-jcp-1994} that imposes the pCCD wavefunction as the fixed reference function.~\cite{frozen-pccd-jcp-2014,pccd-tailored-jctc-2022}
In general, a tCC wave function has the ansatz
%%%%%%%%%%%%%%%%%%%%%%%%%%%%%%%%%%%%%%%%%%%%%%%%%%%%%%%%%%%%%%%%%%%%%%%%%%%%%%%
%%%%%%%%%%%%%%%%%%%%%%%%%%%%%%%%%%%%%%%%%%%%%%%%%%%%%%%%%%%%%%%%%%%%%%%%%%%%%%%
\begin{equation}\label{eq:tcc}
\ket{{\rm tCC}} = e^{\hat{T}} \ket{\Phi_0} = e^{\hat{T}^{\rm ext}} e^{\hat{T}^{\rm int}} \ket{\Phi_0},
\end{equation}
%%%%%%%%%%%%%%%%%%%%%%%%%%%%%%%%%%%%%%%%%%%%%%%%%%%%%%%%%%%%%%%%%%%%%%%%%%%%%%%
%%%%%%%%%%%%%%%%%%%%%%%%%%%%%%%%%%%%%%%%%%%%%%%%%%%%%%%%%%%%%%%%%%%%%%%%%%%%%%%
where $\ket{\Phi_0}$ is a reference Slater determinant and $\hat{T}$ is the cluster operator that is partitioned into a sum of two disjoint cluster operators, $\hat{T}^{\rm int}$ and $\hat{T}^{\rm ext}$.
By construction, $\hat{T}^{\rm int}$ and $\hat{T}^{\rm ext}$ commute if the cluster operators are pure particle-hole excitation operators and the above equation is valid.
In pCCD-tCC,~\cite{frozen-pccd-jcp-2014,pccd-tailored-jctc-2022,geminal-review-pccp-2022} the cluster operator $\hat{T}^{\rm int}$ in eq.~\eqref{eq:tcc} is taken as the pCCD pair-excitation-only cluster operator (cf. eq.~\eqref{eq:pccd}).
Since pCCD describes only electron-pair excitations, the cluster amplitudes in the tCC ansatz are thus divided into the conventional electron-pair amplitudes ($\hat{T}^{\rm int}$) and non-pair (or broken-pair) amplitudes ($\hat{T}^{\rm ext}$).
The pCCD-tCC ansatz, therefore, reads
\begin{equation}
\ket{\mathrm{pCCD-tCC}} = e^{\hat{T}^{\rm ext}} \ket{{\rm pCCD}} = e^{\hat{T}^{\rm ext}} e^{\Tp} \ket{\Phi_0}.
\end{equation}
%rom an implementation perspective, the pCCD-tCC model is easily obtained from a conventional CC implementation by \textit{freezing} the pair amplitudes.
%The \textit{freezing} step requires the pCCD electron-pair amplitudes as input data and neglects those pair contributions to the CC vector function (setting them to zero).
%Thus, pCCD-tCC is commonly known as the frozen-pair (fp)CC ansatz.~\cite{frozen-pccd-jcp-2014}
In the current work, the cluster operator of fpCCD is defined as $\hat{T}^{\rm ext} = \hat{T}_2^\prime = \hat T_2 - \Tp$, while the cluster operator of fpCCSD includes also single excitations, $\hat{T}^{\rm ext} = \hat{T}_1 + \hat{T}_2^\prime$.
Using the CC spin-free notation all excitation operators are written in terms of singlet excitation operators $\hat E_a^i$,
%%%%%%%%%%%%%%%%%%%%%%%%%%%%%%%%%%%%%%%%%%%%%%%%%%%%%%%%%%%%%%%%%%%%%%%%%%%%%%%
%%%%%%%%%%%%%%%%%%%%%%%%%%%%%%%%%%%%%%%%%%%%%%%%%%%%%%%%%%%%%%%%%%%%%%%%%%%%%%%
\begin{equation}
\label{eq:singlet_excitation_op}
    \hat E_a^i = \hat a_{a}^\dagger \hat a_{i} + \hat a_{\bar a}^\dagger \hat a_{\bar i}.
\end{equation}
%%%%%%%%%%%%%%%%%%%%%%%%%%%%%%%%%%%%%%%%%%%%%%%%%%%%%%%%%%%%%%%%%%%%%%%%%%%%%%%
%%%%%%%%%%%%%%%%%%%%%%%%%%%%%%%%%%%%%%%%%%%%%%%%%%%%%%%%%%%%%%%%%%%%%%%%%%%%%%%
For spin-free double excitations, the $\hat T_2$ cluster operator takes on the form
%%%%%%%%%%%%%%%%%%%%%%%%%%%%%%%%%%%%%%%%%%%%%%%%%%%%%%%%%%%%%%%%%%%%%%%%%%%%%%%
%%%%%%%%%%%%%%%%%%%%%%%%%%%%%%%%%%%%%%%%%%%%%%%%%%%%%%%%%%%%%%%%%%%%%%%%%%%%%%%
\begin{equation}
    \hat T_2 = \frac{1}{2} \sum_{i j a b} t_{ij}^{ab} \, \hat E_{a}^{i} \hat E_{b}^{j},
    \label{eq:ccd_spinadapted}
\end{equation}
%%%%%%%%%%%%%%%%%%%%%%%%%%%%%%%%%%%%%%%%%%%%%%%%%%%%%%%%%%%%%%%%%%%%%%%%%%%%%%%
%%%%%%%%%%%%%%%%%%%%%%%%%%%%%%%%%%%%%%%%%%%%%%%%%%%%%%%%%%%%%%%%%%%%%%%%%%%%%%%
while single excitations reduce to
%%%%%%%%%%%%%%%%%%%%%%%%%%%%%%%%%%%%%%%%%%%%%%%%%%%%%%%%%%%%%%%%%%%%%%%%%%%%%%%
%%%%%%%%%%%%%%%%%%%%%%%%%%%%%%%%%%%%%%%%%%%%%%%%%%%%%%%%%%%%%%%%%%%%%%%%%%%%%%%
\begin{equation}
    \hat T_1 = \sum_{i a} t_{i}^{a} \, \hat E_{a}^{i}.
    \label{eq:ccs_spinadapted}
\end{equation}
%%%%%%%%%%%%%%%%%%%%%%%%%%%%%%%%%%%%%%%%%%%%%%%%%%%%%%%%%%%%%%%%%%%%%%%%%%%%%%%
%%%%%%%%%%%%%%%%%%%%%%%%%%%%%%%%%%%%%%%%%%%%%%%%%%%%%%%%%%%%%%%%%%%%%%%%%%%%%%%
The correlation energy of the fpCC ansatz can be obtained by solving the corresponding time-independent projected Schr\"odinger equation
%%%%%%%%%%%%%%%%%%%%%%%%%%%%%%%%%%%%%%%%%%%%%%%%%%%%%%%%%%%%%%%%%%%%%%%%%%%%%%%
%%%%%%%%%%%%%%%%%%%%%%%%%%%%%%%%%%%%%%%%%%%%%%%%%%%%%%%%%%%%%%%%%%%%%%%%%%%%%%%
\begin{equation}\label{eq:fpcc-se}
    \bra{K} e^{-\Tp} e^{-\hat T^{\rm ext}} \hat{H}_N e^{\hat{T}^{\rm ext}} e^{\Tp} \ket{\Phi_0} = 0,
\end{equation}
%%%%%%%%%%%%%%%%%%%%%%%%%%%%%%%%%%%%%%%%%%%%%%%%%%%%%%%%%%%%%%%%%%%%%%%%%%%%%%%
%%%%%%%%%%%%%%%%%%%%%%%%%%%%%%%%%%%%%%%%%%%%%%%%%%%%%%%%%%%%%%%%%%%%%%%%%%%%%%%
where $\bra{K}$ are all singly, doubly-, etc.~excited determinants that can be generated by the external excitation operator $\hat{T}^{\rm ext}$ acting on the pCCD reference state $\ket{\Phi_0}$ and $\hat{H}_N $ is the molecular Hamiltonian in its normal-product form.
Thus, we can define the fpCC-similarity transformed Hamiltonian in normal-product form as~\cite{eom-fpccsd-faraday-disccus-2026}
%%%%%%%%%%%%%%%%%%%%%%%%%%%%%%%%%%%%%%%%%%%%%%%%%%%%%%%%%%%%%%%%%%%%%%%%%%%%%%%
%%%%%%%%%%%%%%%%%%%%%%%%%%%%%%%%%%%%%%%%%%%%%%%%%%%%%%%%%%%%%%%%%%%%%%%%%%%%%%%
\begin{equation}\label{eq:fpcc-effham}
    \hat{\mathcal{H}}_N^{\mathrm{(fpCC)}}
    = e^{-\Tp} e^{-\hat T^{\rm ext}} \hat{H}_N e^{\hat{T}^{\rm ext}} e^{\Tp}.
\end{equation}
We should stress that we work within a spin-free picture, where all excitation operators and projection manifolds are spin-summed and hence all excitation operators correspond to singlet spin-free excitations.

%%%%%%%%%%%%%%%%%%%%%%%%%%%%%%%%%%%%%%%%%%%%%%%%%%%%%%%%%%%%%%%%%%%%%%%%%%%%%%%
%%%%%%%%%%%%%%%%%%%%%%%%%%%%%%%%%%%%%%%%%%%%%%%%%%%%%%%%%%%%%%%%%%%%%%%%%%%%%%%
\subsection{EOM-fpCCD and EOM-ptCCSD excited states models.}
%%%%%%%%%%%%%%%%%%%%%%%%%%%%%%%%%%%%%%%%%%%%%%%%%%%%%%%%%%%%%%%%%%%%%%%%%%%%%%%
%%%%%%%%%%%%%%%%%%%%%%%%%%%%%%%%%%%%%%%%%%%%%%%%%%%%%%%%%%%%%%%%%%%%%%%%%%%%%%%
Within the EOM formalism,~\cite{ eom-cc-rowe-rmp-1968, eom-cc-cpl-1989, eom-ccsd-bartlett-jcp-1993} excited states are computed using a linear CI-type ansatz,
\begin{equation}\label{eq:eom-ci}
        \hat{R} = \sum_\mu c_\mu \hat{\tau}_\mu.
\end{equation}
In the above equation, the sum runs over all (spin-free) excitations present in the cluster operator as well as the identity operator $\hat{\tau}_0$.
The latter accounts for for the ground-state contributions to the targeted excited state.
The excited state is computed by $\hat{R}$ acting on the CC reference state,
\begin{equation}
        \ket{\Psi} = \hat{R} e^{\Tt} \ket{\Phi_0} = \sum_\mu {c_\mu \hat{\tau}_\mu} e^{\Tt} \ket{\Phi_0}.
\end{equation}
The above ansatz is used to derive the well-known EOM-CCSD equations,
\begin{equation}\label{eq:eomcc}
        [\hat{\mathcal{H}}_N^{(\rm CC)},\hat{R}] \ket{\Phi_0}  = \omega \hat{R} \ket{\Phi_0},
\end{equation}
containing the similarity-transformed Hamiltonian in normal-product form, and the excitation energies $\omega=(E-E_0)$ with respect to the CC ground state.
The corresponding EOM-CCSD excited states and excitation energies are obtained by diagonalizing the $\hat{\mathcal{H}}_N^{(\rm CC)}$ represented in the singly (S) and doubly (D) excited  configurational space.
If we treat the $\hat{\mathcal{H}}_N^{\mathrm{(fpCC)}}$ Hamiltonian defined by eq.~\ref{eq:fpcc-effham} within the tailored CC picture, the corresponding EOM diagonalization problem requires the addition of pair (P) excited states in the configuration space,~\cite{pccd-tailored-bartlett-jcp-2023}
\begin{widetext}
\begin{equation}\label{eq:hptcc}
\bm H^{\rm EOM-ptCCSD} =
        \begin{bmatrix}                                                         
          0         & \bra{0} \Hfp \ket{S} & \bra{0} \Hfp \ket{P} & \bra{0} \Hfp \ket{D} \\
          0         & \bra{S} \Hfp \ket{S} & \bra{S} \Hfp \ket{P} & \bra{S} \Hfp \ket{D} \\
\bra{P}\Hfp\ket{0}  & \bra{P} \Hfp \ket{S} & \bra{P} \Hfp \ket{P} & \bra{P} \Hfp \ket{D} \\
          0         & \bra{D} \Hfp \ket{S} & \bra{D} \Hfp \ket{P} & \bra{D} \Hfp \ket{D} \\                      
        \end{bmatrix},
\end{equation}
\end{widetext}
as the pair-amplitude equations are not satisfied for $\hat{\mathcal{H}}_N^{\mathrm{(fpCC)}}$, that is, $\bra{P}\hat{\mathcal{H}}_N^{\mathrm{(fpCC)}}\ket{0}] \neq 0$.
That model is named EOM pair tailored CCSD (EOM-ptCCSD).~\cite{pccd-tailored-bartlett-jcp-2023}
Unlike the standard EOM-CCSD approach, in the EOM-ptCCSD model a non-vanishing term in the first column arises as the electron-pair excitation amplitudes do not satisfy the projected Schr\"odinger equation (the $\Tp$ amplitudes satisfy only $\bra{P}\Hp\ket{0} = 0$).
This issue can be resolved if $\Hn$ in the diagonalization problem is adjusted so that pair and broken-pair amplitudes are treated differently, that is, if the seniority sectors for seniority-zero and seniority-non-zero states are decoupled in both the ground and excited-state model.

If we enforce the projection equations to be satisfied for each excitation manifold (or seniority sector), the Hamiltonian matrix to be diagonalized becomes 
\begin{widetext}
\begin{equation}\label{eq:hfpcc}
\bm H^{\rm EOM-fpCCSD} =
        \begin{bmatrix}                                                         
          0         & \bra{0} \Hfp \ket{S} & \bra{0} \Hfp \ket{P} & \bra{0} \Hfp \ket{D}  \\
          0         & \bra{S} \Hfp \ket{S} & \bra{S} \Hfp \ket{P} & \bra{S} \Hfp \ket{D} \\
          0         & \bra{P} \Hp  \ket{S} & \bra{P} \Hp  \ket{P} & \bra{P} \Hp  \ket{D} \\
          0         & \bra{D} \Hfp \ket{S} & \bra{D} \Hfp \ket{P} & \bra{D} \Hfp \ket{D} \\                       
        \end{bmatrix}.                                            
\end{equation}
\end{widetext}

Thus, the first column ``vanishes'' as the pCCD amplitude equations ($\bra{P}\Hp\ket{0} = 0$) are satisfied, while all broken-pair amplitudes are optimized within the fpCC step.
The corresponding frozen pair model is called EOM-fpCCSD.~\cite{eom-fpccsd-faraday-disccus-2026}

%%%%%%%%%%%%%%%%%%%%%%%%%%%%%%%%%%%%%%%%%%%%%%%%%%%%%%%%%%%%%%%%%%%%%%%%%%%%%%%
%%%%%%%%%%%%%%%%%%%%%%%%%%%%%%%%%%%%%%%%%%%%%%%%%%%%%%%%%%%%%%%%%%%%%%%%%%%%%%%
\subsection{Approximate transition moments within the EOM-fpCCSD and EOM-ptCCSD formalism.}
%%%%%%%%%%%%%%%%%%%%%%%%%%%%%%%%%%%%%%%%%%%%%%%%%%%%%%%%%%%%%%%%%%%%%%%%%%%%%%%
%%%%%%%%%%%%%%%%%%%%%%%%%%%%%%%%%%%%%%%%%%%%%%%%%%%%%%%%%%%%%%%%%%%%%%%%%%%%%%%   
Computing excited-state properties within the EOM-CCSD/EOM-fpCCSD/EOM-ptCCSD framework requires the evaluation of both right ($R_k$) and left ($L_k$) eigenvectors of a given excited-state root $N$. 
While the $R_k$ are sufficient to obtain excitation energies, the non-Hermitian nature of the EOM similarity-transformed Hamiltonian makes the $L_k$  and $R_k$ eigenvectors distinct. Consequently, both sets of vectors are essential for the proper calculation of transition dipole moments, oscillator strengths, and other one-electron excited-state properties.
To circumvent the computationally expensive evaluation of the left eigenvectors $L_k$, we employ the matrix inverse approximation,~\cite{approximate-left-eom-eigenvectors-jcp-2011,eom-ccsd-expectation-value-tdms-jcp-2017}
\begin{equation}
\bm L^{\dagger} \approx (\bm R^{\dagger} \bm R)^{-1} \bm R,
\end{equation}
which by construction satisfies the biorthonormality condition for the corresponding right eigenvectors.~\cite{eom-ccsd-expectation-value-tdms-jcp-2017}
In the above equation, $\bm L^\dagger$ and $\bm R$ are $N\times N_{\rm root}$ matrices, where $N$ is the number of (spin-free) excitations and $N_{\rm root}$ the number of targeted excited states.

Furthermore, we use the expectation value approximation to avoid additional computational overhead from solving the CC $\Lambda$ equations,~\cite{xcc-ijqc-1993, td-xcc-collect-czech-chem-commun-2005, korona-s-operator-cc-jcp-2006,approximate-lambda-cc-jcp-2011,tdm-xcc-jcp-2014, tdm-xcc-jcp-2017}
\begin{equation}
\hat{\Lambda}^\dagger \approx \hat{T}.
\end{equation}
The expectation value approach has recently been explored for pCCD-based ground-state properties, yielding very satisfactory results.~\cite{pccd-expectation-value-1dm-jpca-2025, pccd-in-pccd-jcp-2026} 

Based on these two approximations, we can compute the left and right transition dipole moments, 
\begin{align}
T^R_{0k} &= \mel{0}{(1+ \hat{\Lambda})\,\bar{D}\,R_k}{0},\\
T^L_{k0} &= \mel{0}{L_k\,\bar{D}}{0},
\end{align}
where $\bar{D}$ denotes the similarity transformed dipole moment operator. 

The exact expressions for the elements of the EOM-CC transition density matrices (employed to evaluate $T^L_{k0}$ and $T^R_{0k}$) for the spin-free case are provided in the Appendix.
These are then used to compute the transition properties between the ground and $k$-th excited state,~\cite{eom-ccsd-1-rdm-jcp-1993, eom-ccsd-1-rdm-jctc-2019} like the dipole strength (DS),
\begin{equation}
\textrm{DS}=\abs{T_{k}}^{2} = T^L_{k0}T^R_{0k}
=
\mel{0}{L_k\,\bar{D}}{0}
\mel{0}{(1+\Lambda)\,\bar{D}\,R_k}{0}
\end{equation}
or the experimentally observable oscillator strengths (OS),
\begin{equation}
{\rm OS}=\frac{2}{3}\omega_k \abs{T_{k}}^{2}.
\end{equation}

%%%%%%%%%%%%%%%%%%%%%%%%%%%%%%%%%%%%%%%%%%%%%%%%%%%%
%%%%%%        Comput Details
%%%%%%%%%%%%%%%%%%%%%%%%%%%%%%%%%%%%%%%%%%%%%%%%%%%%
\section{Computational details} \label{sec:comput-det}
%%%%%%%%%%%%%%%%%%%%%%%%%%%%%%%%%%%%%%%%%%%%%%%%%%%%
%%%%%%%%%%%%%%%%%%%%%%%%%%%%%%%%%%%%%%%%%%%%%%%%%%%%
All calculations were performed using a developer version of the PyBEST software package (v2.2.0dev0).~\cite{pybest-paper-cpc-2021, pybest-paper-update1-cpc-2024, pybest-gpu-jctc-2024}
% Electron-repulsion integrals were evaluated via Cholesky decomposition.~\cite{cholesky-koch-jcp-2003, cholesky-review-2011, cholesky-vesality-wires-2024}
The cc-pVXZ (X = D, T, Q) and the aug-cc-pVTZ basis sets were employed.~\cite{cc-pvxz-dunning-jcp-1989, aug-cc-pvxz-jcp-1992}
In the correlated calculations, PyBEST’s default frozen-core approximation was applied (1s for C and O).
For each system, the twelve lowest-lying roots were computed in the EOM-CC calculations, consistent with the LR-CCSD reference.~\cite{lr-pccd-jctc-2024}
In our calculations, two sets of orbitals were used: the canonical HF orbitals and natural pCCD orbitals.
The choice of molecular orbital basis in the CCSD/EOM-CCSD workflow is indicated in parentheses, like EOM-CCSD(HF) or EOM-CCSD(pCCD). 
The Cholesky representation of the electron repulsion integrals~\cite{cholesky-koch-jcp-2003, cholesky-review-2011, cholesky-vesality-wires-2024} (threshold set to $10^{-5}$) was used for the computations of furan with aug-cc-pVTZ and cc-pVQZ basis sets.
The XYZ structures of the investigated compounds were taken from Ref.~\citenum{lr-pccd-jctc-2024} and are provided in the Supporting Information.
%%%%%%%%%%%%%%%%%%%%%%%%%%%%%%%%%%%%%%%%%%%%%%%%%%%%
%%%%%%%%%%%%%%%%%%%%%%%%%%%%%%%%%%%%%%%%%%%%%%%%%%%%
\begin{figure*}[t]
    \centering
     \includegraphics[width=0.85\linewidth]{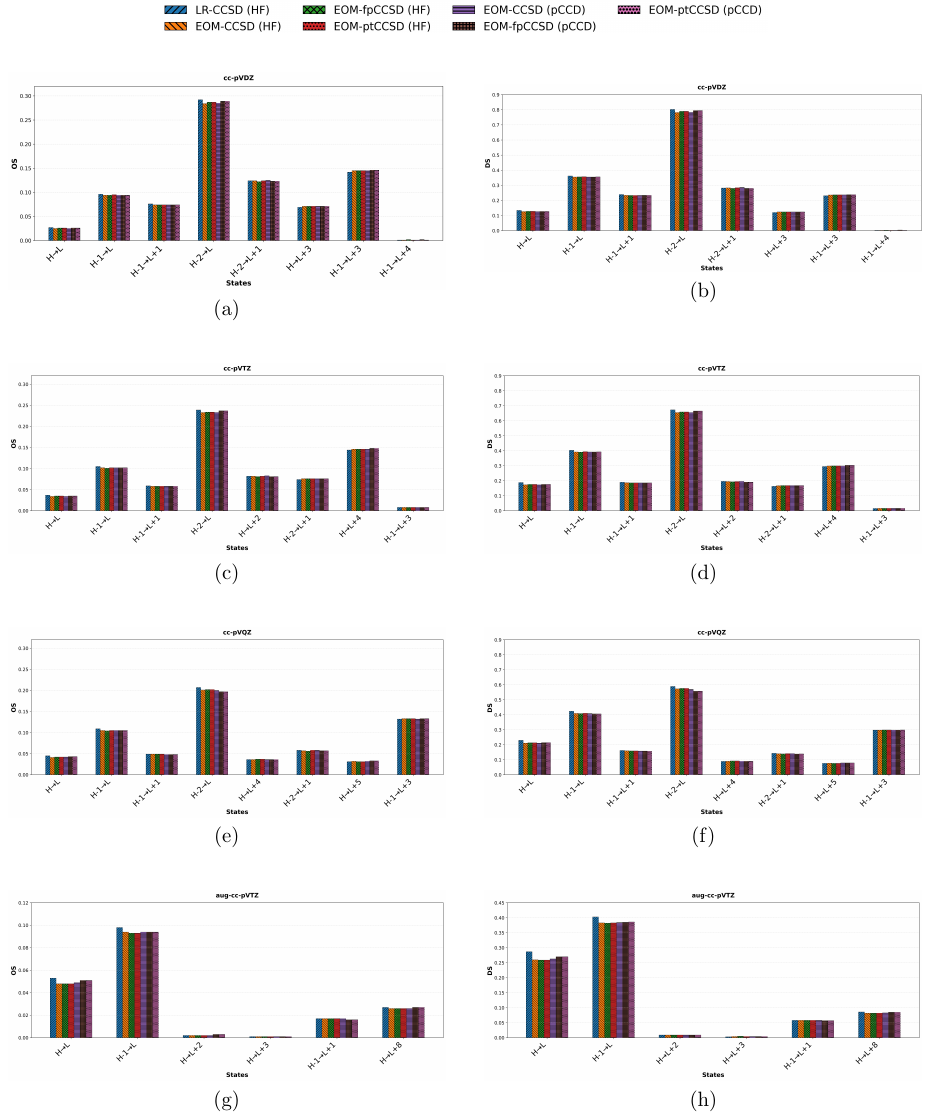}
    \caption{Oscillator strengths (OSs) and Dipole Strengths (DS) for the lowest-lying excited states of the water molecule.
    The LR-CCSD(HF), EOM-CCSD(HF/pCCD), EOM-fpCCSD(HF/pCCD), and EOM-ptCCSD(HF/pCCD) results are obtained with the cc-pVDZ, cc-pVTZ, and cc-pVQZ basis sets. 
    HF and pCCD in parentheses denote calculations using canonical Hartree–Fock and natural pCCD orbitals, respectively. 
    H and L at the bottom of each subfigure label the HOMO and LUMO orbitals.}
    \label{fig:tdm-water}
\end{figure*}
%%%%%%%%%%%%%%%%%%%%%%%%%%%%%%%%%%%%%%%%%%%%%%%%%%%%
%%%%%%%%%%%%%%%%%%%%%%%%%%%%%%%%%%%%%%%%%%%%%%%%%%%%
\begin{figure*}[t]
    \centering
     \includegraphics[width=0.85\linewidth]{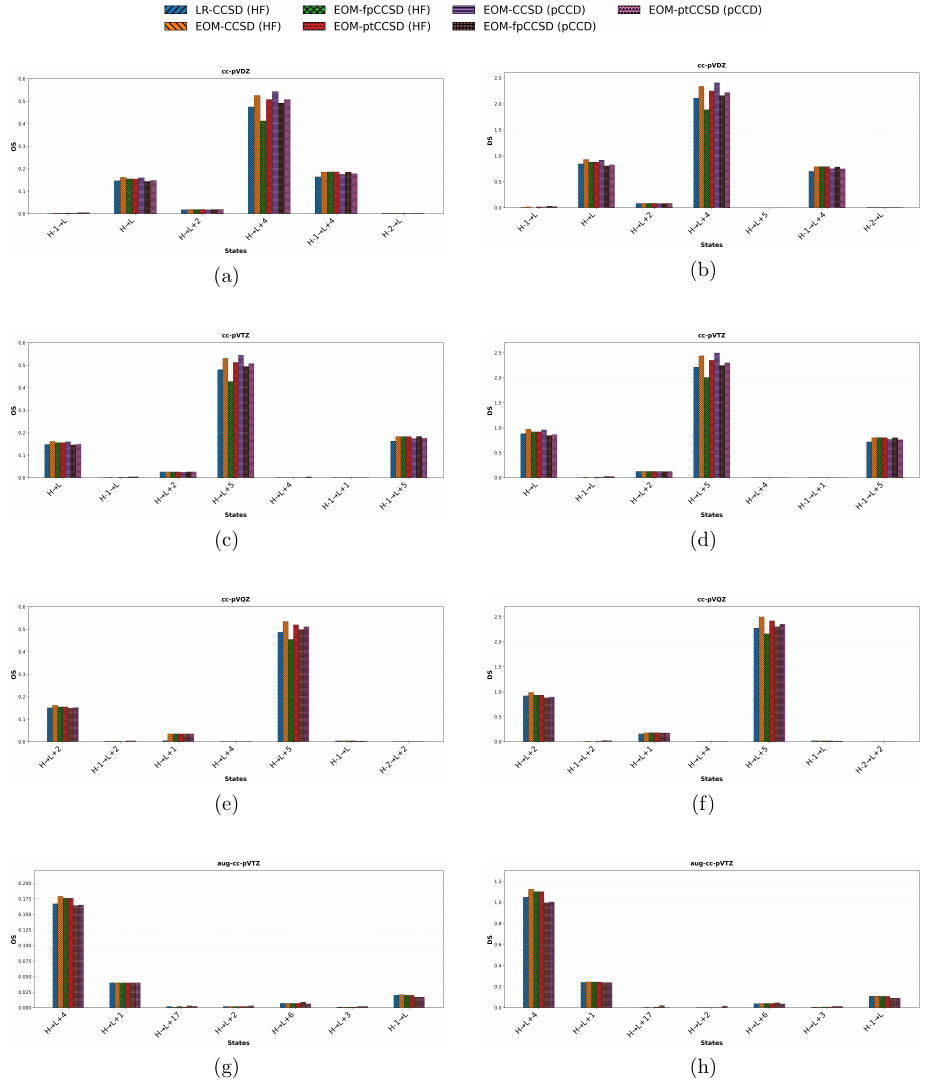}
    \caption{Oscillator strengths (OSs) and Dipole Strengths (DS) for the low-lying excited states of the furan molecule.
    The LR-CCSD(HF), EOM-CCSD(HF/pCCD), EOM-fpCCSD(HF/pCCD), and EOM-ptCCSD(HF/pCCD) results are obtained with the cc-pVDZ, cc-pVTZ, and cc-pVQZ basis sets. 
    HF and pCCD in parentheses denote calculations using canonical Hartree–Fock and natural pCCD orbitals, respectively. 
    H and L at the bottom of each subfigure label the HOMO and LUMO orbitals.}
    \label{fig:tdm-furan}
\end{figure*}
%%%%%%%%%%%%%%%%%%%%%%%%%%%%%%%%%%%%%%%%%%%%%%%%%%%%
%%%%%%%%%%%%%%%%%%%%%%%%%%%%%%%%%%%%%%%%%%%%%%%%%%%%
%%%%%%%%%%%%%%%%%%%%%%%%%%%%%%%%%%%%%%%%%%%%%%%%%%%%
%%%%%%        Results
%%%%%%%%%%%%%%%%%%%%%%%%%%%%%%%%%%%%%%%%%%%%%%%%%%%%
\section{Results and discussion}\label{sec:results}
%%%%%%%%%%%%%%%%%%%%%%%%%%%%%%%%%%%%%%%%%%%%%%%%%%%%
%%%%%%%%%%%%%%%%%%%%%%%%%%%%%%%%%%%%%%%%%%%%%%%%%%%%
%\subsection{Vertical excitation energies.}
We test our methods for the low-lying vertical excited-state properties of water and furan for various basis set sizes.
Specifically, we benchmark EOM-CCSD(HF/pCCD), EOM-fpCCSD(HF/pCCD), and EOM-ptCCSD(HF/pCCD) against the LR-CCSD(HF) reference using the cc-pVDZ, cc-pVTZ, cc-pVQZ, and aug-cc-pVTZ basis sets. 
Our approximate excited-state properties are not size-intensive by design, unlike LR-CCSD.~\cite{size-intensivity-eom-lr-jcp-1994} 
However, the size-intensity problem is negligible for the investigated molecules as LR-CCSD and a standard EOM-CCSD give identical excited state properties.~\cite{lr-pccd-jctc-2024} 

Our results are summarized in Figures~\ref{fig:tdm-water} and~\ref{fig:tdm-furan}. 
Those Figures illustrate the performance of the EOM-CC models for the water and furan molecules, respectively.
In both figures, the left column displays the OSs, while the right column shows the DSs for the bright states across the cc-pVDZ, cc-pVTZ, cc-pVQZ, and aug-cc-pVTZ basis sets.
We note that the excited-state ordering is determined using the EOM-CCSD(HF) results, and the character of the corresponding states in EOM-fpCCSD and EOM-ptCCSD is assigned accordingly.
The dominant orbital contributions are expressed in terms of the positions of the Highest Occupied Molecular Orbital (HOMO) and the Lowest Unoccupied Molecular Orbital (LUMO) in the HF ordering. 
The statistical performance of each method across different basis sets is presented in Tables~\ref{tab:ave_statistics_hf} and~\ref{tab:ave_statistics_pccd} for HF and pCCD orbitals, respectively. 
Dark (optically forbidden) excited states are omitted from Figures~\ref{fig:tdm-water} and~\ref{fig:tdm-furan} and the statistical analysis, but are reported in the SI.

Figure~\ref{fig:tdm-water} illustrates the performance of the EOM-CC models for the water molecule.
All methods perform similarly well, accurately reproducing the trends in both OSs and DSs.
Differences between the cc-pVDZ, cc-pVTZ, and cc-pVQZ basis sets are small across all methods.
The aug-cc-pVTZ basis set exhibits somewhat distinct features in the excited states of water; however, all methods still perform very similarly in this basis.
For the furan molecule (Figure~\ref{fig:tdm-furan}), the differences between basis sets are a bit more pronounced.
In particular, for the most intense peaks (those with the largest OSs and DSs), the spread among the EOM-CC methods increases.
A combined statistical analysis depicted in Table~\ref{tab:overall_statistics_water_furan} shows that the EOM-fpCCSD approach yields the smallest percentage errors relative to the LR-CCSD reference data when using both canonical HF and natural pCCD orbitals.
The EOM-ptCCSD method also outperforms the standard EOM-CCSD approach.
The larger discrepancies in the OSs across these methods stem from differences in excitation energies arising from the non-negligible admixture of doubly excited states relative to LR-EOM-CCSD. 

In summary, the numerical results demonstrate that the choice of orbital basis (HF versus pCCD) has a negligible influence on the excited-state description within the investigated EOM-CCSD, EOM-fpCCSD, and EOM-ptCCSD models.
That opens the way to modeling excited-state properties in complex systems where the HF reference is not stable or reliable.
This behavior contrasts with the much simpler LR-pCCD model.\cite{lr-pccd-jctc-2024}
Most importantly, both EOM-fpCCSD and EOM-ptCCSD yield a significantly improved description of excited-state properties compared to the standard EOM-CCSD method.

%%%%%%%%%%%%%%%%%%%%%Table 1  %%%%%%%%%%%%%%%%%%%%%%%%%%%%%%
\begin{table}[htbp]
\centering
\caption{Average statistical errors of Dipole Strengths (DS) and oscillator strengths (OS) for water and furan using canonical HF molecular orbitals. 
ME, MAE, SD, RMSE, and MPE denote the mean error, mean absolute error, standard deviation, root mean square error, and mean percentage error, respectively. }
\scriptsize
\setlength{\tabcolsep}{2pt}
\begin{tabular}{lcccccc}
\hline
 & \multicolumn{2}{c}{EOM-CCSD} &
   \multicolumn{2}{c}{EOM-fpCCSD} &
   \multicolumn{2}{c}{EOM-ptCCSD} \\

 & DS & OS & DS & OS & DS & OS \\
\hline

\multicolumn{7}{c}{\textbf{cc-pVDZ}}\\
\hline
ME     & 0.019 & 0.004 & $-$0.006 & $-$0.002 & 0.013 & $-$0.003 \\
MAE    & 0.022 & 0.005 & 0.006 & 0.002 & 0.015 & 0.003 \\
SD     & 0.031 & 0.007 & 0.006 & 0.002 & 0.021 & 0.005 \\
RMSE   & 0.029 & 0.006 & 0.008 & 0.002 & 0.019 & 0.005 \\
MPE    & 6.2   & 6.0   & 1.9   & 2.4   & 4.1   & 4.2   \\
\hline

\multicolumn{7}{c}{\textbf{cc-pVTZ}}\\
\hline
ME     & 0.018 & 0.004 & $-$0.006 & $-$0.002 & 0.012 & $-$0.003 \\
MAE    & 0.022 & 0.005 & 0.006 & 0.002 & 0.015 & 0.003 \\
SD     & 0.032 & 0.007 & 0.003 & 0.001 & 0.021 & 0.005 \\
RMSE   & 0.029 & 0.006 & 0.007 & 0.002 & 0.019 & 0.004 \\
MPE    & 6.1   & 5.8   & 2.0   & 2.2   & 4.1   & 4.2   \\
\hline

\multicolumn{7}{c}{\textbf{cc-pVQZ}}\\
\hline
ME     & 0.013 & 0.004 & $-$0.006 & $-$0.001 & 0.007 & $-$0.003 \\
MAE    & 0.019 & 0.005 & 0.006 & 0.001 & 0.012 & 0.004 \\
SD     & 0.027 & 0.007 & 0.002 & 0.001 & 0.016 & 0.006 \\
RMSE   & 0.023 & 0.006 & 0.007 & 0.001 & 0.014 & 0.005 \\
MPE    & 6.1   & 8.1   & 2.4   & 1.3   & 3.9   & 6.1   \\
\hline

\multicolumn{7}{c}{\textbf{aug-cc-pVTZ}}\\
\hline
ME     & 0.002 & 0.000 & 0.000 & 0.000 & 0.001 & 0.000 \\
MAE    & 0.007 & 0.001 & 0.005 & 0.001 & 0.006 & 0.001 \\
SD     & 0.010 & 0.002 & 0.008 & 0.002 & 0.008 & 0.002 \\
RMSE   & 0.007 & 0.001 & 0.006 & 0.001 & 0.006 & 0.001 \\
MPE    & 6.0   & 5.7   & 5.1   & 5.0   & 5.2   & 5.2   \\
\hline

\end{tabular}
\label{tab:ave_statistics_hf}
\end{table}

%%%%%%%%%%%End of Table 1 %%%%%%%%%
% Table 2 %%%%%%%%%%%%%%%%%%%%%%%%%%%%%%%%
\begin{table}[htbp]
\centering
\caption{Average statistical errors of Dipole Strengths (DS) and oscillator strengths (OS) for water and furan using pCCD-optimized natural orbitals.
ME, MAE, SD, RMSE, and MPE denote the mean error, mean absolute error, standard deviation, root mean square error, and mean percentage error, respectively.}
\scriptsize
\setlength{\tabcolsep}{2pt}
\begin{tabular}{lcccccc}
\hline
 & \multicolumn{2}{c}{EOM-CCSD} &
   \multicolumn{2}{c}{EOM-fpCCSD} &
   \multicolumn{2}{c}{EOM-ptCCSD} \\

 & DS & OS & DS & OS & DS & OS \\
\hline

\multicolumn{7}{c}{\textbf{cc-pVDZ}}\\
\hline
ME     & 0.020 & 0.004 & 0.005 & 0.002 & 0.007 & $-$0.003 \\
MAE    & 0.023 & 0.005 & 0.007 & 0.002 & 0.009 & 0.003 \\
SD     & 0.032 & 0.007 & 0.009 & 0.003 & 0.012 & 0.004 \\
RMSE   & 0.030 & 0.007 & 0.008 & 0.003 & 0.011 & 0.004 \\
MPE    & 6.4   & 6.2   & 2.0   & 2.5   & 2.5   & 3.3   \\
\hline

\multicolumn{7}{c}{\textbf{cc-pVTZ}}\\
\hline
ME     & 0.018 & 0.004 & 0.003 & $-$0.006 & 0.005 & $-$0.002 \\
MAE    & 0.023 & 0.005 & 0.006 & 0.006 & 0.008 & 0.002 \\
SD     & 0.032 & 0.007 & 0.009 & 0.008 & 0.011 & 0.004 \\
RMSE   & 0.029 & 0.006 & 0.007 & 0.008 & 0.009 & 0.003 \\
MPE    & 6.1   & 5.9   & 1.9   & 7.0   & 2.2   & 3.1   \\
\hline

\multicolumn{7}{c}{\textbf{cc-pVQZ}}\\
\hline
ME     & n.c. & n.c. & $-$0.002 & 0.001 & 0.001 & $-$0.002 \\
MAE    & n.c. & n.c. & 0.005 & 0.003 & 0.008 & 0.004 \\
SD     & n.c. & n.c. & 0.007 & 0.004 & 0.011 & 0.005 \\
RMSE   & n.c. & n.c. & 0.005 & 0.003 & 0.008 & 0.004 \\
MPE    & n.c. & n.c. & 2.1   & 4.7   & 2.9   & 5.7   \\
\hline

\multicolumn{7}{c}{\textbf{aug-cc-pVTZ}}\\
\hline
ME     & n.c. & n.c. & $-$0.004 & 0.000 & $-$0.004 & 0.000 \\
MAE    & n.c. & n.c. & 0.004 & 0.000 & 0.004 & 0.000 \\
SD     & n.c. & n.c. & 0.000 & 0.000 & 0.001 & 0.000 \\
SD-pop & n.c. & n.c. & 0.000 & 0.000 & 0.001 & 0.000 \\
RMSE   & n.c. & n.c. & 0.004 & 0.001 & 0.004 & 0.001 \\
MPE    & n.c. & n.c. & 3.6   & 2.0   & 3.8   & 2.2   \\
\hline
\end{tabular}
\footnotetext[1]{n.c. = not computed due to convergence difficulties }
\label{tab:ave_statistics_pccd}
\end{table}
%%%%%%%%%%%End of Table 2 %%%%%%%%%

\begin{table}[htbp]
\centering
\caption{Average statistical errors of transition dipole moments (DS) and oscillator strengths (OS) for water and furan relative to LR-CCSD reference values using four different basis sets (averaged over cc-pVDZ, cc-pVTZ, cc-pVQZ, and aug-cc-pVTZ).
ME, MAE, SD, RMSE, and MPE denote the mean error, mean absolute error, standard deviation, root mean square error, and mean percentage error, respectively.}
\scriptsize
\setlength{\tabcolsep}{3pt}
\begin{tabular}{llcccccc}
\hline
 &  &
\multicolumn{2}{c}{EOM-CCSD} &
\multicolumn{2}{c}{EOM-fpCCSD} &
\multicolumn{2}{c}{EOM-ptCCSD} \\
&
& DS & OS & DS & OS & DS & OS \\
\hline

\multicolumn{8}{c}{\textbf{Water (HF orbitals)}}\\
\hline
ME   & & $-$0.005 & $-$0.001 & $-$0.004 & $-$0.001 & $-$0.004 & 0.001 \\
MAE  & &  0.005 &  0.001 &  0.004 &  0.001 &  0.004 & 0.001 \\
SD   & &  0.001 &  0.000 &  0.001 &  0.000 &  0.001 & 0.000 \\
RMSE & &  0.005 &  0.001 &  0.004 &  0.001 &  0.004 & 0.001 \\
MPE  & &  3.1   &  2.5   &  2.8   &  2.4   &  2.7   & 2.2   \\
\hline

\multicolumn{8}{c}{\textbf{Water (pCCD orbitals)}}\\
\hline
ME   & & $-$0.004 & $-$0.001 & $-$0.004 & $-$0.001 & $-$0.004 & 0.001 \\
MAE  & &  0.004 &  0.001 &  0.004 &  0.001 &  0.004 & 0.001 \\
SD   & &  0.001 &  0.000 &  0.002 &  0.001 &  0.002 & 0.001 \\
RMSE & &  0.004 &  0.001 &  0.004 &  0.001 &  0.004 & 0.001 \\
MPE  & &  2.9   &  2.3   &  2.5   &  1.7   &  2.4   & 1.7   \\
\hline

\multicolumn{8}{c}{\textbf{Furan (HF orbitals)}}\\
\hline
ME   & &  0.031 &  0.007 & $-$0.005 & $-$0.001 &  0.020 & -0.005 \\
MAE  & &  0.031 &  0.007 &  0.008 &  0.002 &  0.020 &  0.005 \\
SD   & &  0.015 &  0.004 &  0.007 &  0.002 &  0.010 &  0.003 \\
RMSE & &  0.033 &  0.008 &  0.008 &  0.002 &  0.021 &  0.006 \\
MPE  & &  9.2   & 10.3   &  2.8   &  3.0   &  6.0   &  7.6   \\
\hline

\multicolumn{8}{c}{\textbf{Furan (pCCD orbitals)}}\\
\hline
ME   & & n.c. & n.c. &  0.005 & $-$0.001 &  0.008 & -0.004 \\
MAE  & & n.c. & n.c. &  0.007 &  0.005 &  0.010 &  0.004 \\
SD   & & n.c. & n.c. &  0.007 &  0.007 &  0.009 &  0.003 \\
RMSE & & n.c. & n.c. &  0.008 &  0.006 &  0.011 &  0.004 \\
MPE  & & n.c. & n.c. &  2.3   &  6.4   &  3.3   &  5.4   \\
\hline
\end{tabular}
\footnotetext[1]{n.c. = not computed due to convergence difficulties }
\label{tab:overall_statistics_water_furan}
\end{table}
%%%%%%%%%%%%%%%%%%%%%%%%%%%%%%%%%%%%%%%%%%%%%%%%%%%%
%%%%%%        Conclusions
%%%%%%%%%%%%%%%%%%%%%%%%%%%%%%%%%%%%%%%%%%%%%%%%%%%%
\section{Conclusions}\label{sec:conclusions}
%%%%%%%%%%%%%%%%%%%%%%%%%%%%%%%%%%%%%%%%%%%%%%%%%%%%
%%%%%%%%%%%%%%%%%%%%%%%%%%%%%%%%%%%%%%%%%%%%%%%%%%%%
In this work, we implemented and benchmarked a simple computational scheme to obtain TDMs and OSs from various equation-of-motion coupled-cluster methods: EOM-CCSD, EOM-fpCCSD, and EOM-ptCCSD.
These quantities are computed using both (delocalized) canonical HF and (localized) natural pCCD orbitals and are benchmarked against LR-CCSD reference data.~\cite{lr-pccd-jctc-2024}
Our approach has a lower computational cost than traditional LR-CCSD as the solution of the CC $\Lambda$ equations and left eigenvectors of the EOM-CC calculation are avoided.
However, the resulting excited-state properties are not size-intensive by design, due to the presence of disconnected terms in the right transition density matrices.~\cite{size-intensivity-eom-lr-jcp-1994, eom-ccsd-1-rdm-jctc-2019}

Benchmark calculations on the low-lying vertical excitation energies, TDMs, and OSs of water and furan in various basis set sizes (including the augmented sets) yield highly encouraging results.
Specifically, the computed TDMs and OSs are in very good agreement with the LR-CCSD reference values.
Overall errors in the TDMs and OSs remain below 5\% in most cases and show no significant dependence on basis set size or the inclusion of diffuse functions.
Importantly, both canonical HF and natural pCCD orbitals produce comparable accuracy when paired with the EOM-CCSD, EOM-fpCCSD, and EOM-ptCCSD methods.
These findings demonstrate that the recently introduced single-reference coupled-cluster approaches~\cite{eom-fpccsd-faraday-disccus-2026} can be reliably extended to study excited-state spectra and properties of complex electronic structures, including charge-transfer states and states with substantial doubly excited character.

%%%%%%%%%%%%%%%%%%%%%%%%%%%%%%%%%%%%%%%%%%%%%%%%%%%%
%%%%%%     Acknowledgment   
%%%%%%%%%%%%%%%%%%%%%%%%%%%%%%%%%%%%%%%%%%%%%%%%%%%%
\section*{Acknowledgment}
%%%%%%%%%%%%%%%%%%%%%%%%%%%%%%%%%%%%%%%%%%%%%%%%%%%%
%%%%%%%%%%%%%%%%%%%%%%%%%%%%%%%%%%%%%%%%%%%%%%%%%%%%
S.~J., and P.~T.~acknowledge financial support from the SONATA BIS research grant from the National Science Centre, Poland (Grant No. 2021/42/E/ST4/00302). 
Funded/Co-funded by the European Union (ERC, DRESSED-pCCD, 101077420).
Views and opinions expressed are, however, those of the author(s) only and do not necessarily reflect those of the European Union or the European Research Council. Neither the European Union nor the granting authority can be held responsible for them.

%%%%%%%%%%%%%%%%%%%%%%%%%%%%%%%%%%%%%%%%%%%%%%%%%%%%
%%%%%%     Appendix   
%%%%%%%%%%%%%%%%%%%%%%%%%%%%%%%%%%%%%%%%%%%%%%%%%%%%
\appendix
\renewcommand{\theequation}{A.\arabic{equation}}
\setcounter{equation}{0}

\section{Transition Density Matrices and Dipole Moments}

All transition density matrices are given for restricted spatial orbitals.
Only the $\alpha$ spin component is provided, as $\rho^\alpha = \rho^\beta$.
All equations are given for the spin-free case, where all CC amplitudes, $\Lambda$ amplitudes, and left and right eigenvectors correspond to the spin-free case.

In the following, $l^{p\ldots}_{q\ldots}$ indicates the elements of the left eigenvector $L_k$, $r^{p\ldots}_{q\ldots}$ encodes the elements of the right eigenvector $R_k$, $\lambda^{p\ldots}_{q\ldots}$ are the elements of the $\Lambda$ amplitudes, and $t^{p\ldots}_{q\ldots}$ are the CC amplitudes.
For better readability, we drop the state-dependence $k$ in all transition density matrix equations, while the subscript $k$ is used as a summation index.
We employ the standard convention of summation indices, where $i,j,k,m,\ldots$ run over occupied and $a,b,c,d,\ldots$ over virtual (molecular) orbital indices.

The elements of the (spin-free) left transition density matrix read
%%%%%%%%%%%%%%%%%%%%%%%%%%%%%%%%%%%%%%%%%%%%%%%%%%%%
\begin{equation}
    (\rho^L_{k0})_{ij}^{\alpha} = \delta_{ij}l_0 - \frac{1}{2}\sum_{c}l_j^ct_i^c-\frac{1}{2}\sum_{kcd} l_{kj}^{cd} t_{ki}^{cd}
 \end{equation}
%%%%%%%%%%%%%%%%%%%%%%%%%%%%%%%%%%%%%%%%%%%%%%%%%%%%
\begin{equation}
    (\rho^L_{k0})_{ab}^{\alpha} =\frac{1}{2}\sum_{k}l_k^at_k^b +\frac{1}{2}\sum_{kmc} l_{km}^{ac} t_{km}^{bc}
\end{equation}
%%%%%%%%%%%%%%%%%%%%%%%%%%%%%%%%%%%%%%%%%%%%%%%%%%%%
\begin{equation}
\begin{aligned}
    (\rho^L_{k0})_{ia}^{\alpha} &= 
    l_0t_i^a + \sum_{kc}l_k^c t_{ik}^{ac}
    -\frac{1}{2}\sum_{kc} l_{k}^{c} t_{ki}^{ac}
    - \frac{1}{2}\sum_{kc} l_{k}^{c} t_{k}^{a}t_{i}^{c} 
    \\
    &\phantom{=}
    - \frac{1}{2}\sum_{kmcd} l_{km}^{cd} t_{ki}^{cd} t_m^a 
    -\frac{1}{2}\sum_{kmcd}l_{km}^{cd} t_{km}^{ca} t_i^d 
\end{aligned}
\end{equation}
%%%%%%%%%%%%%%%%%%%%%%%%%%%%%%%%%%%%%%%%%%%%%%%%%%%%
\begin{equation}
    (\rho^L_{k0})_{ai}^{\alpha} =\frac{1}{2}l_i^a
\end{equation}

The equations for the right transition density matrix are not fully connected and contain disconnected contributions.
The elements of the (spin-free) right transition density matrix read 
%%%%%%%%%%%%%%%%%%%%%%%%%%%%%%%%%%%%%%%%%%%%%%%%%%%
\begin{equation}
\begin{aligned}
    (\rho^R_{0k})_{ij}^{\alpha} &=
    \delta_{ij}r_0 
    -\frac{1}{2}\sum_c \lambda_j^c t_i^c r_0 
    -\frac{1}{2}\sum_c \lambda_j^c r_i^c 
    \\
    &\phantom{=}
    -\frac{1}{2}\sum_{kcd} \lambda_{jk}^{dc} t_{ki}^{cd} r_0 
    -\frac{1}{2}\sum_{kcd} \lambda_{jk}^{dc} t_{i}^{d} r_k^c    
    \\
    &\phantom{=}
    -\frac{1}{2}\sum_{kcd} \lambda_{jk}^{dc} r_{ki}^{cd} 
\end{aligned}
\end{equation}
%%%%%%%%%%%%%%%%%%%%%%%%%%%%%%%%%%%%%%%%%%%%%%%%%%%%
\begin{equation}
\begin{aligned}
    (\rho^R_{0k})_{ia}^{\alpha} &=
    t_{i}^{a}r_0  + r_i^a
    -\frac{1}{2}\sum_{kc} \lambda_{k}^{c} t_i^c r_{k}^{a} 
    -\frac{1}{2}\sum_{kc} \lambda_{k}^{c} t_{k}^{a} r_{i}^{c} 
    +\sum_{kc} \lambda_{k}^{c} t_{i}^{a} r_{k}^{c} 
    \\
    &\phantom{=}
    +\sum_{kmcd} \lambda_{km}^{cd} t_{ik}^{ac} r_{m}^{d} 
    -\frac{1}{2}\sum_{kmcd} \lambda_{km}^{cd} t_{ki}^{ac} r_{m}^{d}
    -\frac{1}{2}\sum_{kmcd} \lambda_{km}^{cd} t_{ik}^{dc} r_{m}^{a}
    \\
    &\phantom{=}
    -\frac{1}{2}\sum_{kmcd} \lambda_{km}^{cd} t_{mk}^{ac} r_{i}^{d}
    +\frac{1}{2}\sum_{kmcd} \lambda_{km}^{cd} t_{i}^{a} r_{km}^{cd}
    -\frac{1}{2}\sum_{kmcd} \lambda_{km}^{cd} t_{m}^{a} r_{ik}^{dc}
    \\
    &\phantom{=}
    -\frac{1}{2}\sum_{kmcd} \lambda_{km}^{cd} t_{i}^{d} r_{mk}^{ac}
    +\sum_{kc} \lambda_{k}^{c} t_{ki}^{ca} r_0
    -\frac{1}{2}\sum_{kc} \lambda_{k}^{c} t_{ik}^{ca} r_0
    \\
    &\phantom{=}
    -\frac{1}{2}\sum_{kc} \lambda_{k}^{c} t_{i}^{c} t_{k}^{a} r_0
    -\frac{1}{2}\sum_{kmcd} \lambda_{km}^{cd} t_{ik}^{dc} t_{m}^{a} r_0
    -\frac{1}{2}\sum_{kmcd} \lambda_{km}^{cd} t_{mk}^{ac} t_{i}^{d} r_0
    \\
    &\phantom{=}
    +\sum_{kc} \lambda_{k}^{c} r_{ki}^{ca}
    -\frac{1}{2}\sum_{kc} \lambda_{k}^{c} r_{ik}^{ca}
    -\frac{1}{2}\sum_{kmcd} \lambda_{km}^{cd} t_{i}^{c} t_{k}^{a} r_{m}^{d}
    \end{aligned}
\end{equation}
%%%%%%%%%%%%%%%%%%%%%%%%%%%%%%%%%%%%%%%%%%%%%%%%%%%%
\begin{equation}
    (\rho^R_{0k})_{ai}^{\alpha} =
    \frac{1}{2} \lambda_i^a r_0 
    +\frac{1}{2}\sum_{kc} \lambda_{ki}^{ca} r_k^c 
\end{equation}
%%%%%%%%%%%%%%%%%%%%%%%%%%%%%%%%%%%%%%%%%%%%%%%%%%%%
\begin{equation}
\begin{aligned}
    (\rho^R_{0k})_{ab}^{\alpha} &=
    \frac{1}{2} \sum_k \lambda_k^a t_k^b r_0 
    +\frac{1}{2}\sum_{k} \lambda_{k}^{a} r_k^b 
    \\
    &\phantom{=}
    +\frac{1}{2}\sum_{kmc} \lambda_{km}^{ca} t_{km}^{cb} r_0 
    +\frac{1}{2}\sum_{kmc} \lambda_{km}^{ca} t_{m}^b r_k^c 
    \\
    &\phantom{=}
    +\frac{1}{2}\sum_{kmc} \lambda_{km}^{ca} r_{km}^{cb} 
\end{aligned}
\end{equation}

The corresponding transition dipole moments are then evaluated using the dipole integrals $\{d_{pq}\}$ (expressed in the molecular orbital basis)
%%%%%%%%%%%%%%%%%%%%%%%%%%%%%%%%%%%%%%%%%%%%%%%%%%%%
\begin{equation}
    T^L_{k0} = 2\sum_{pq} (\rho^L_{k0})_{pq}^\alpha d_{pq}
\end{equation}
%%%%%%%%%%%%%%%%%%%%%%%%%%%%%%%%%%%%%%%%%%%%%%%%%%%%
and
%%%%%%%%%%%%%%%%%%%%%%%%%%%%%%%%%%%%%%%%%%%%%%%%%%%%
\begin{equation}
    T^R_{0k} = 2\sum_{pq} (\rho^R_{0k})_{pq}^\alpha d_{pq}.
\end{equation}
%%%%%%%%%%%%%%%%%%%%%%%%%%%%%%%%%%%%%%%%%%%%%%%%%%%%

%%%%%%%%%%%%%%%%%%%%%%%%%%%%%%%%%%%%%%%%%%%%%%%%%%%%
%%%%%%     Bibliography
%%%%%%%%%%%%%%%%%%%%%%%%%%%%%%%%%%%%%%%%%%%%%%%%%%%%
\bibliography{p}
\bibliographystyle{apsrev4-1}
%%%%%%%%%%%%%%%%%%%%%%%%%%%%%%%%%%%%%%%%%%%%%%%%%%%%
%%%%%%%%%%%%%%%%%%%%%%%%%%%%%%%%%%%%%%%%%%%%%%%%%%%%

\end{document}